\title{Tests of Pythia8 Rescattering Model}
\author{
        Tasnuva Chowdhury \\
                Shahjalal University of Science and Technology\\
       Sylhet, Bangladesh.\\
            \and
       Deepak Kar\\
        University of Witwatersrand\\
        Johannesburg, South Africa.\\
}
\date{\today}

\documentclass[12pt]{article}

\usepackage[update,prepend]{epstopdf} 
\usepackage{cite,mcite}
\usepackage{rotating}
\usepackage{graphicx}
\usepackage{relsize}
\usepackage {placeins}
\usepackage{mathptmx}
\usepackage{xspace}
\usepackage{xcolor}
\usepackage{longtable}
\usepackage{booktabs}
\usepackage{multicol}
\usepackage{multirow}
\usepackage[small,nooneline,center]{subfigure}
\usepackage[italic]{hepparticles}
\usepackage{amssymb}
\usepackage{amsmath}
\usepackage{setspace}
\usepackage{threeparttable}
\usepackage{footnote}
\usepackage{url}
\RequirePackage{tikz}
\usepackage{relsize}

\newcommand{\kbd}[1]{{\smaller{\texttt{#1}}}}
\def\pT{\ensuremath{p_\textrm{T}}\xspace}

\def\ZpT{\ensuremath{p_{\textrm{T}}^{\textrm{Z}}}}

\newcommand{\Z}{{$Z$-boson}\xspace}

\def\pythiaeight{Pythia\,8\xspace}

\newcommand{\FigRef}[1]{Figure~\ref{#1}\xspace}

\begin{document}
\maketitle

\begin{abstract}

One of the most poorly understood phenomenon in hadron collisions, is the so called multiple parton
interaction (MPI). Apart from one quark or gluon each from each colliding
proton, additional quarks or gluons can interact as well, and these
can not be calculated from first principles. The concept of
rescattering has been introduced recently
in \pythiaeight event generator, where particles
originating from these secondary interactions can interact again with
quarks or gluons from incoming protons.
In this paper, we look at events with a $Z$-boson, to find
observables which can potentially be sensitive to this rescattering
effect. 
While jet-balance observables do not show visible
difference, charged particle distributions in different azimuthal
regions show some difference. The parameters controlling
MPI can be \textit{tuned} to give a good description of data
with rescattering.

\end{abstract}

%%%%%%%%%%%%%%%%%%%%%%%%%%%%%%%%%%%%%%%%%%%%%%%%%%%
\section{Introduction}

Monte Carlo event generators are used extensively in collider physics. \pythiaeight is one
of the most commonly used generator~\cite{Sjostrand:2007gs}, using the parton shower approach.
The parton shower approach is based 
on the assumption that a $2\rightarrow n$ process, with a complex final state is achieved by starting from
a simple $2 \rightarrow 2$ process that approximately defines the 
directions and energies of the hardest partons, and adding
a succession of simple parton branchings to build up the full event structure.
The different components of the model are initial and final state radiation (I/FSR), multiple
parton interactions (MPI), fragmentation and hadronisation. In this paper, the focus is on 
MPI.

The collider coordinate system needs to be specified in order to define
the variables. The colliding beams are taken along the $z$-axis, while
the $x$-$y$ plane represents the transverse direction with respect to the beams, 
where the collision output particles are.
Transverse momentum, \pT is defined relative to the beam axis.
The azimuthal angle $\phi$ is measured around the beam axis, and the 
polar angle $\theta$ is measured with respect to~the $z$-axis. 
The pseudorapidity is given by $\eta = -\ln\tan\mspace{-0.1mu}( \theta/2 )$.

The analysis and plots are done using the Rivet~\cite{Buckley:2010ar} analysis framework.

%%%%%%%%%%%%%%%%%%%%%%%%%%%%%%%%%%%%%%%%%%%%%%%%%%%
\section{MPI and Rescattering}

\begin{figure}[h]
  \centering
  \includegraphics[width=0.45\textwidth]{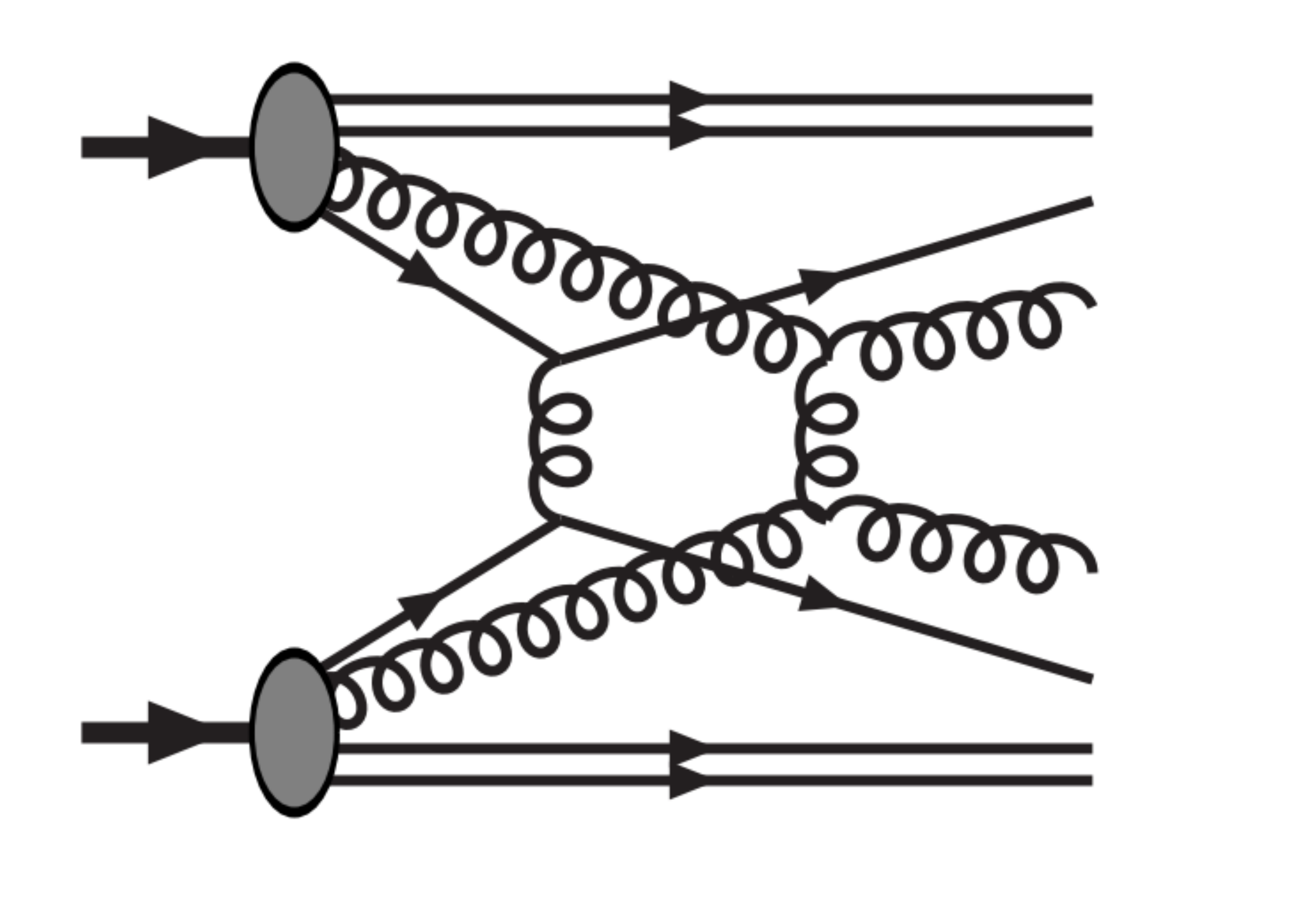}
    \includegraphics[width=0.45\textwidth]{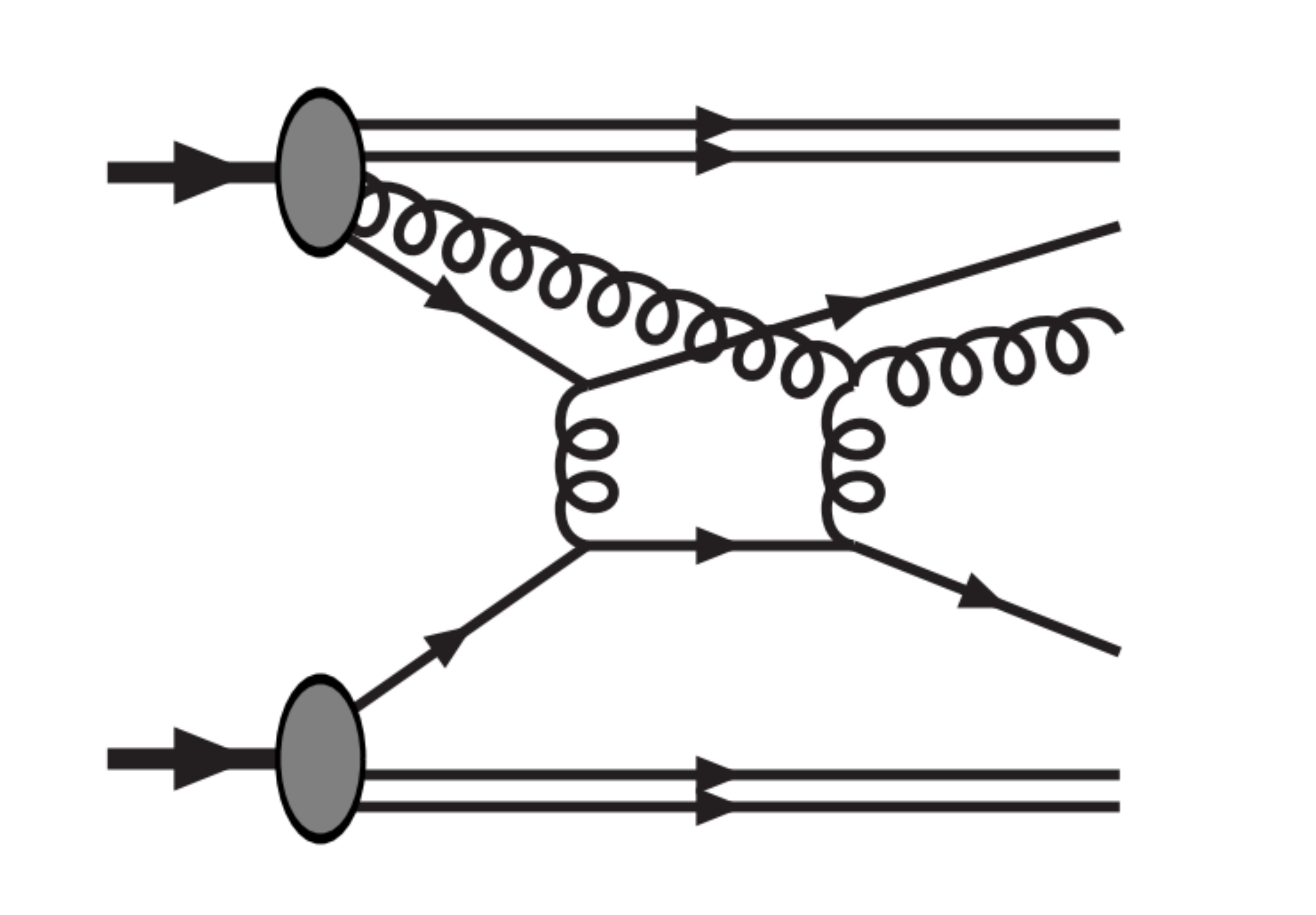}
  \caption{Schematic diagram of MPI and rescattering }
  \label{fig:mpi}
\end{figure}

Since each incoming proton in hadron colliders such as LHC
is a composite object, consisting of many partons, the actual collision
happens between two partons, which is referred to as hard scatter (HS).
However, there exists the possibility of several parton pairs interacting when two hadrons collide, 
which is termed as multiple parton interactions (MPI). Double parton intercation (DPI),
is a special case of MPI, with only one additional scattering. \FigRef{fig:mpi}, left schematically represents
an MPI event,taken from~\cite{Corke:2009tk}.

Among the parameters controlling the strength of MPI in \pythiaeight model, a \pT
cutoff, and $\alpha_s$, the strong coupling constant for MPI will be looked at here.
The \pT cutoff is necessary to regularise the divergence of partonic
interaction cross-section at low-\pT, and larger/smaller values result in less/more MPI.
The $\alpha_s$ value is usually evaluated at the $Z$-mass, and results in
more/less activity from MPI for higher/lower value.

The concept of rescattering~\cite{Corke:2009tk} off MPI has been introduced 
recently in \pythiaeight. This occurs when a parton produced from the HS can 
interact directly with a parton from an incoming proton. This is the simplest example,
and termed single rescattering, as shown in \FigRef{fig:mpi}, right, again taken from~\cite{Corke:2009tk}. 
Double rescattering can also happen, which can be considered as the more general case.

\FloatBarrier

%%%%%%%%%%%%%%%%%%%%%%%%%%%%%%%%%%%%%%%%%%%%%%%%%%%
\section{Setup}

The first step was to check if any kinematic observables can be sensitive to the presence
of rescattering. The \Z production, in association with extra jets was chosen, at it offers
a relatively clean final state. Events with at least $3$ extra jets, each with transverse momentum,
$\pT> 20$ GeV and absolute rapidity, $|y| < 4.4$ were used.

The assumption is that the highest \pT jet (leading jet) will balance the \Z momentum,
and the additional two jets can either originate from ISR, or from MPI. In case they originate
from MPI, they are allowed to undergo rescattering. Now on an individual event-by-event basis
it is unphysical to try to label jets depending on their origin (HS, ISR or MPI), and no attempt
is made to do so. Rather a set of kinematic distributions were looked at, once generating the events
once with rescattering, once without. This is done by:\\
$\kbd{MultipartonInteractions:allowRescatter = on}$\\
$\kbd{MultipartonInteractions:allowDoubleRescatter = on}$\\

The samples were generated with Monash tune~\cite{Skands:2014pea} using
NNPDF2.3LO PDF set~\cite{Carrazza:2013axa}. $5$ million events were generated in each case.

\begin{figure}[tbp]
  \begin{center}
    \begin{tikzpicture}[>=stealth, very thick, scale=1.2]
      %% Make figure text smaller than for the body
      %\smaller
      \small

      %% Circle and region dividers
      \draw[color=blue!80!black] (0, 0) circle (3.0);
      \draw[rotate= 30, color=gray] (-3.0, 0) -- (3.0, 0);
      \draw[rotate=-30, color=gray] (-3.0, 0) -- (3.0, 0);

      %% Delta{phi} definition arcs
      \draw[->, color=black, rotate=-2] (0, 3.5) arc (90:47:3.5) node[right] {$\Delta{\phi}$};
      \draw[->, color=black, rotate=2] (0, 3.5) arc (90:133:3.5) node[left] {$-\Delta{\phi}$};

      %% Tracks
      \draw[->, color=red, ultra thick] (0, 2) -- (0, 4) node[above] {\textcolor{black}{\Z}};
      \draw[->, color=green!70!black, ultra thick] (0, -2) -- ( 0.0, -4);
      \draw[->, color=green!70!black, ultra thick, rotate around={-15:(0,-0.3)}] (0, -2) -- ( 0.0, -4);
      \draw[->, color=green!70!black, ultra thick, rotate around={ 15:(0,-0.3)}] (0, -2) -- ( 0.0, -4);

      %% Region definition labels
      \draw (0,  1.4) node[text width=2cm] {\begin{center} toward \goodbreak $|\Delta\phi| < 60^\circ$ \end{center}};
      \draw (0, -1.1) node[text width=2cm] {\begin{center} away \goodbreak $|\Delta\phi| > 120^\circ$ \end{center}};
      \draw ( 1.7, 0.3) node[text width=3cm] {\begin{center} transverse \goodbreak $60^\circ < |\Delta\phi| < 120^\circ$ \end{center}};
      \draw (-1.7, 0.3) node[text width=3cm] {\begin{center} transverse \goodbreak $60^\circ < |\Delta\phi| < 120^\circ$ \end{center}};
    \end{tikzpicture}
    \caption{Definition of UE regions in the azimuthal angle with respect to the \Z.}
    \label{fig:ueregions}
  \end{center}
\end{figure}
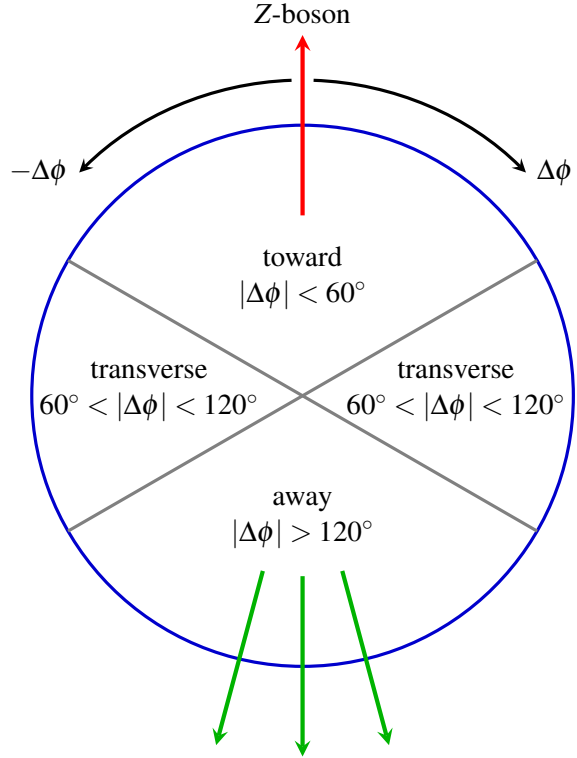

The next step was to compare the simulation to actual data distributions. For this, the published
ATLAS \Z underlying event (UE) analysis~\cite{Aad:2014jgf} was used. In this analysis, the direction of the 
\Z in each event was taken as the direction of HS, and the azimuthal plane was divided into
the toward, transverse and away regions on an event-by-event basis, as shown in \FigRef{fig:ueregions},
Additionally again on an event-by-event basis, the more (less) active transverse side
was labeled as transmax (transmin). 

The transverse region, being perpendicular to the direction of hard scattering, is expected to be most sensitive
to underlying event, which is defined as the additional activity in an event with an identified HS. However,
at the busy LHC environment, the transverse regions receives significant contribution from additional
HS jets, so it can no longer be used as a clean probe of UE. However, 
after the subtraction of the contribution from the
leptons from the \Z decay,
the activity in the toward region is also sensitive to the underlying event.
Since there is no final-state gluon radiation.
It is also expected that transmin region, by construction, will have less contribution 
from HS jets, hence it is also used to probe UE. Following that, the modeling of charged
particle activity in this two regions will be looked at.

\FloatBarrier

%%%%%%%%%%%%%%%%%%%%%%%%%%%%%%%%%%%%%%%%%%%%%%%%%%%
\section{Results}

The kinematic distributions of several variables which are expected to sensitive 
to DPI topology are compared between samples generated with and without rescattering.

The first variable is indicative to \pT balance bewteen hard objects. In the following,
$\vec{\pT^{i}}$, indicates the vector transverse momentum of the $i$th hard object.
Since at least 3 additional jets are required along with the \Z , $i=1$ will denote 
the \ZpT, and $i=2,3,4$ will denote the jets, in order of decreasing \pT.
The value closer to unity denotes the hard objects are more \pT balanced.

\[
\Delta_{ij}^{\pT} = \frac{|\vec{\pT^i} + \vec{\pT^j}|}{ {|\vec{\pT^i}| + |\vec{\pT^j}|}} 
\]

\begin{figure}[h]
  \centering
   \includegraphics[width=0.45\textwidth]{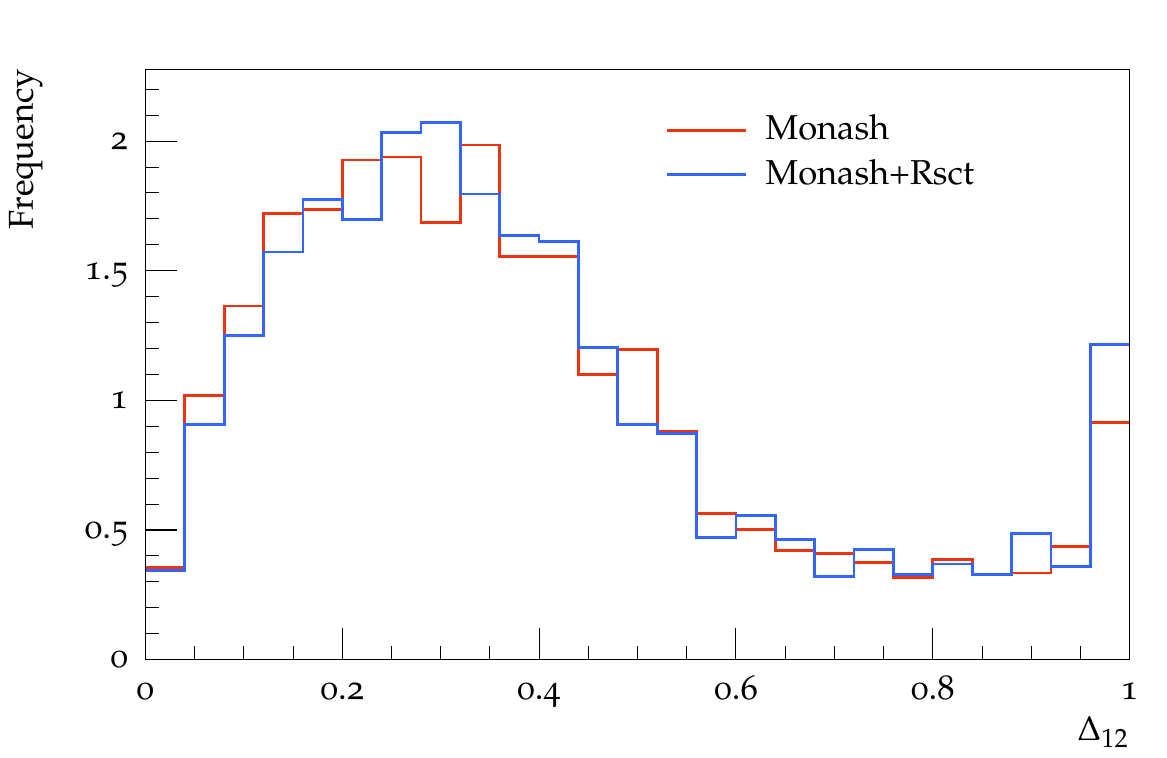}\\[1em]
  \includegraphics[width=0.45\textwidth]{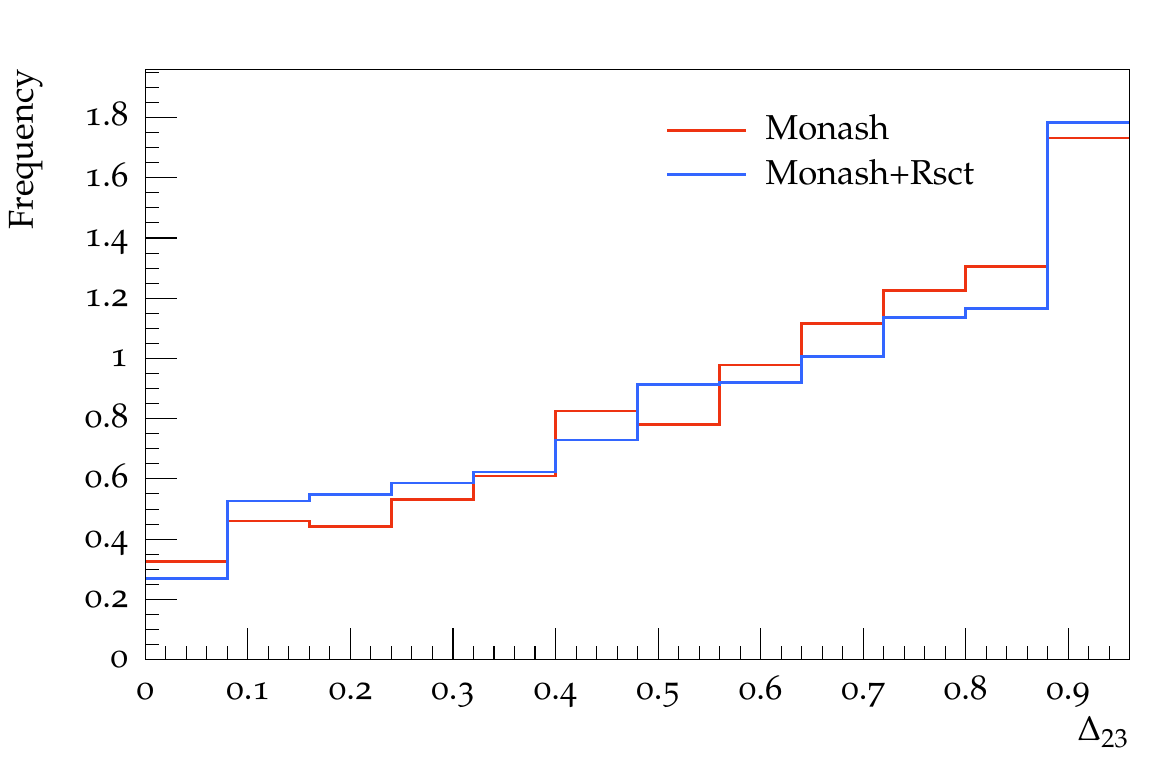}\qquad
  \includegraphics[width=0.45\textwidth]{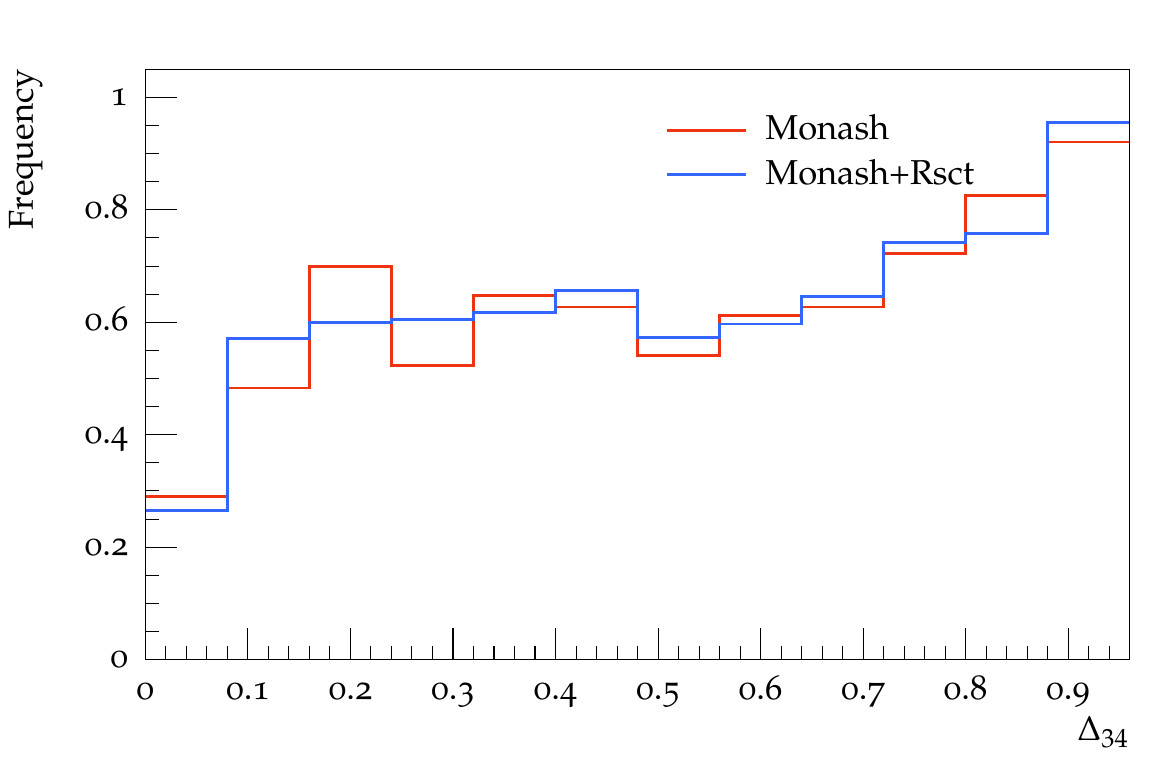}\\[1em]
  \caption{Comparisons between \pythiaeight predictions without and with rescattering for \pT balance observables}
  \label{fig:mc1}
\end{figure}

As \pythiaeight do not produce enough $3$-jet events, the distributions suffer from a lack of statistics, which makes
arriving at any strong conclusions difficult. However, some general trends can be observed. In \FigRef{fig:mc1}, the
$\Delta_{ij}^{\pT}$ variable is can be seen for two combinations. While the difference between events generated
without and with rescattering is not expected to affect the \pT balance between the \Z and the leading jet, 
the 1st and 2nd jet, and 2nd and 3rd jet balances are also fairly insensitive to rescattering.

\begin{figure}[h]
  \centering
  \includegraphics[width=0.45\textwidth]{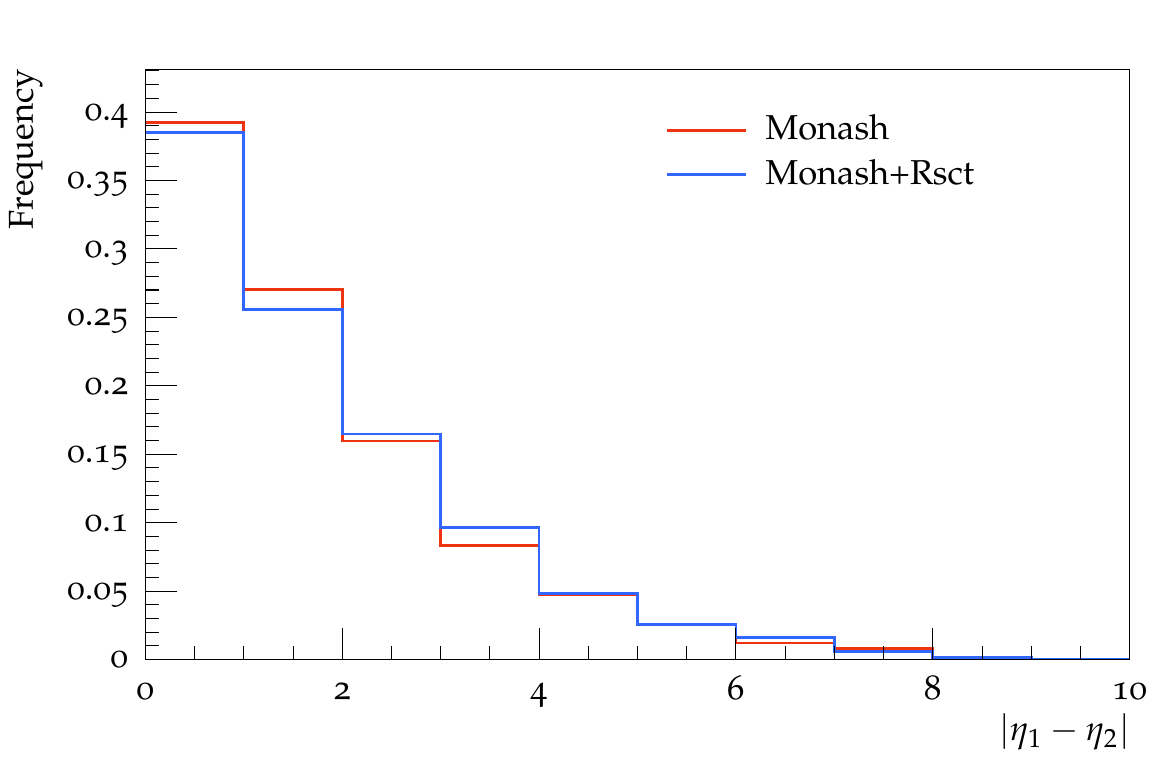}\qquad
  \includegraphics[width=0.45\textwidth]{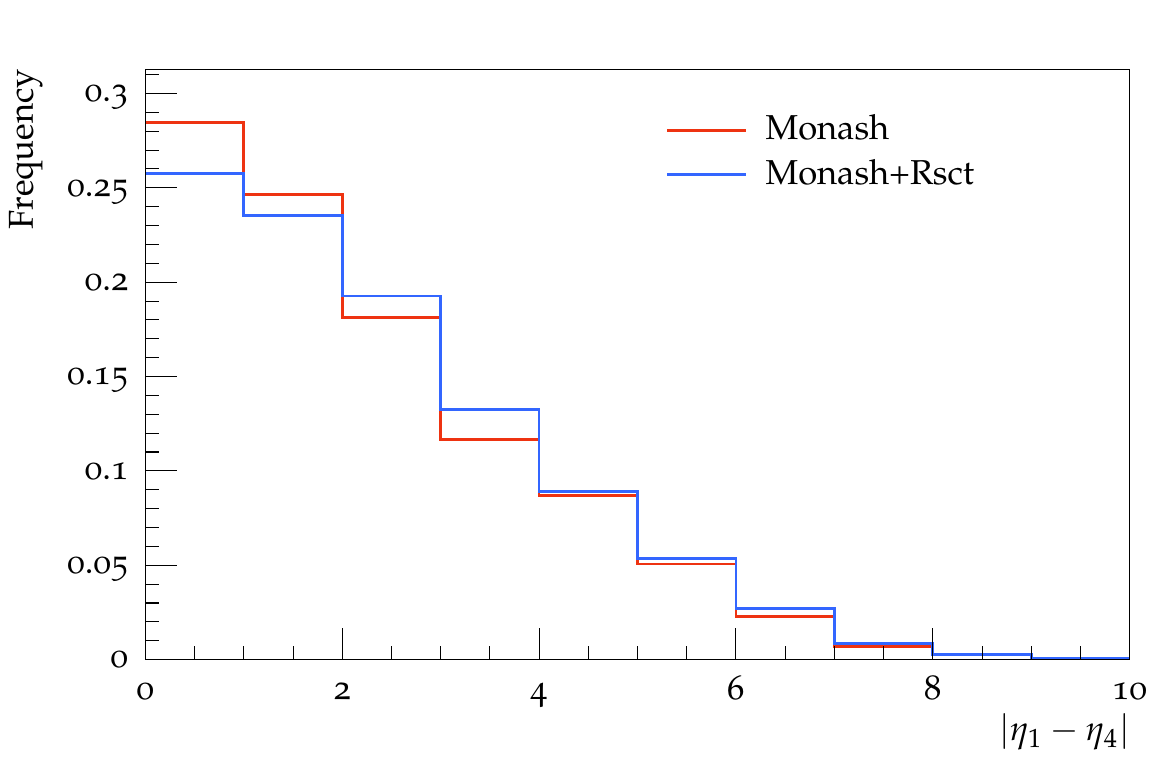}\\[1em]
  \caption{Comparisons between \pythiaeight predictions without and with rescattering for $\eta$ difference observables}
  \label{fig:mc2}
\end{figure}

The angular distributions also suffer from a similar lack of statistics. In \FigRef{fig:mc2}, the $\eta$
differences are looked at, with the indices defined as before. Again the $\eta$ difference between
the \Z and the leading jet is essentially non-existent for events without and with
rescattering, but the $\eta$ difference between the \Z and 3rd leading jet seems to a bit
more when rescattering happens.

\begin{figure}[h]
  \centering
  \includegraphics[width=0.45\textwidth]{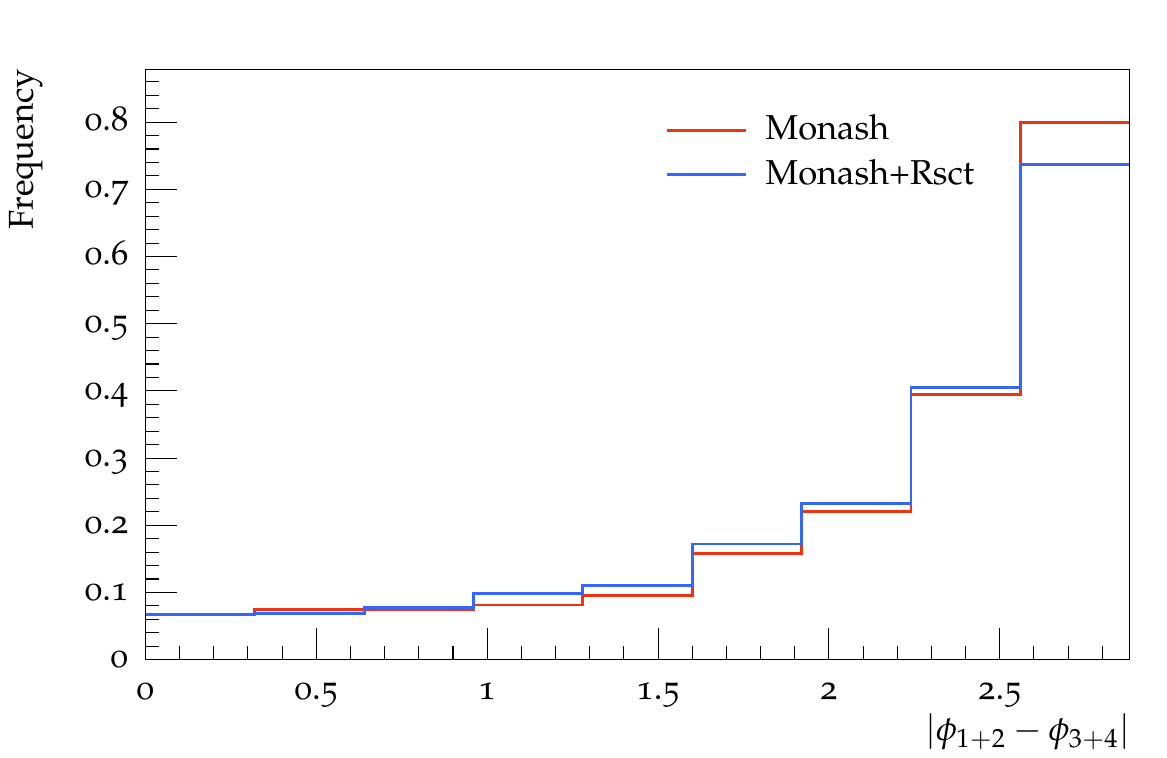}\qquad
  \includegraphics[width=0.45\textwidth]{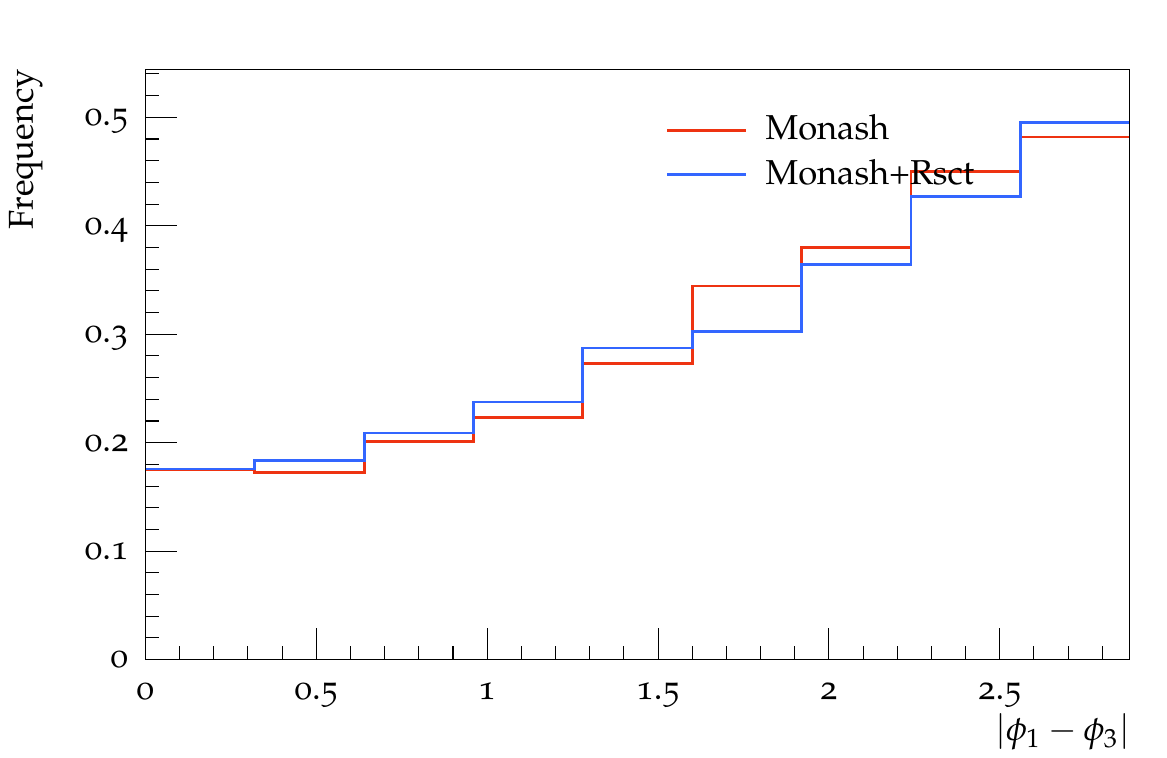}\\[1em]
  \caption{Comparisons between \pythiaeight predictions without and with rescattering for $\phi$ difference observables}
  \label{fig:mc3}
\end{figure}

Finally, in \FigRef{fig:mc3}, the $\phi$ difference is probed, again the indices are defined as before.
The term $\phi_{i+j}$ represents the $\phi$ of resultant four-vector by adding the four-vectors
of hard objects $i$ and $j$. This $\phi$ balance two supposedly balanced pair of hard objects
show hardly any difference, as well as the difference of $\phi$ of the \Z and the 3rd leading jet, for
events without and with rescattering.

\begin{figure}[h]
  \centering
  \includegraphics[width=0.45\textwidth]{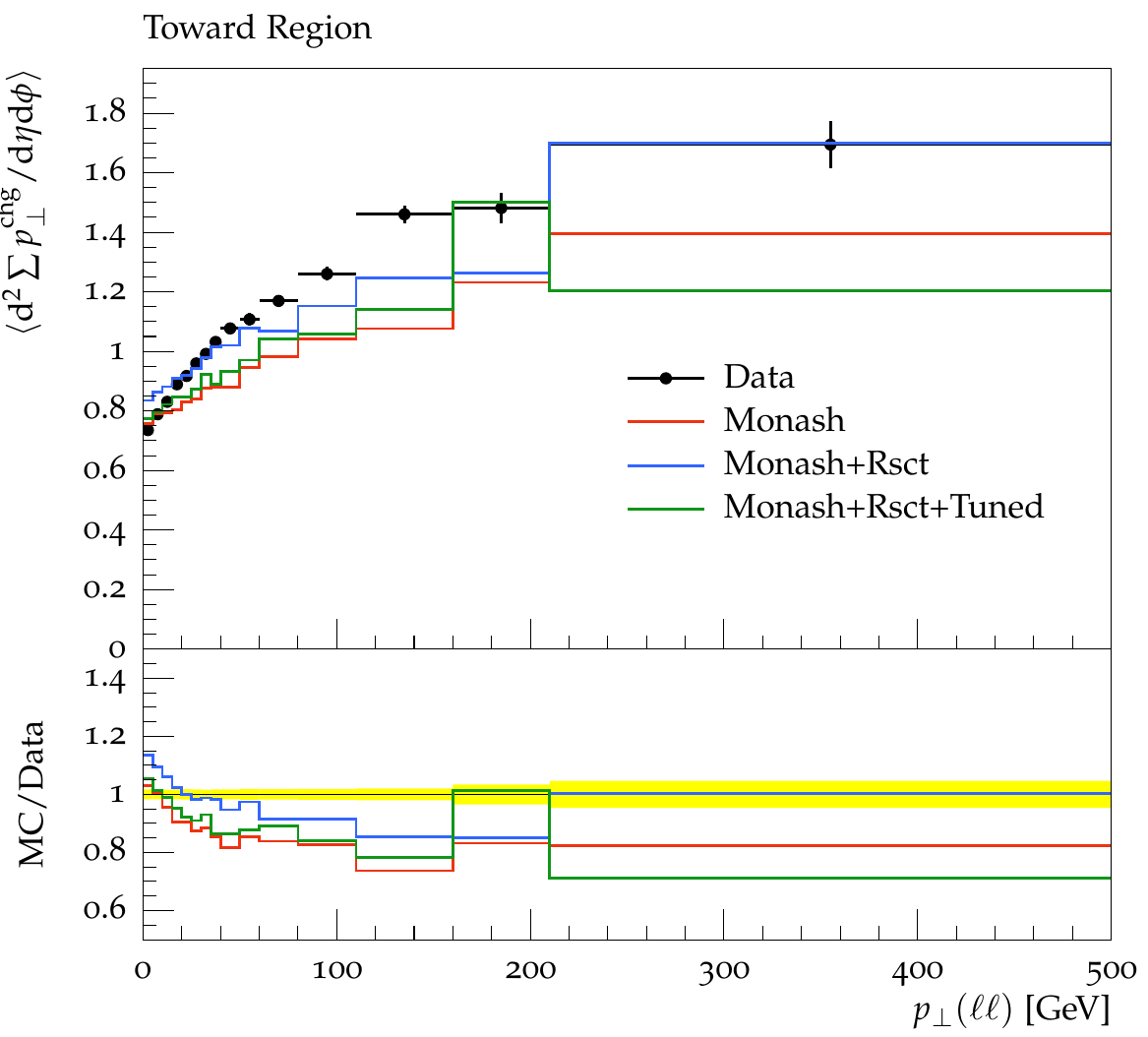}\qquad
  \includegraphics[width=0.45\textwidth]{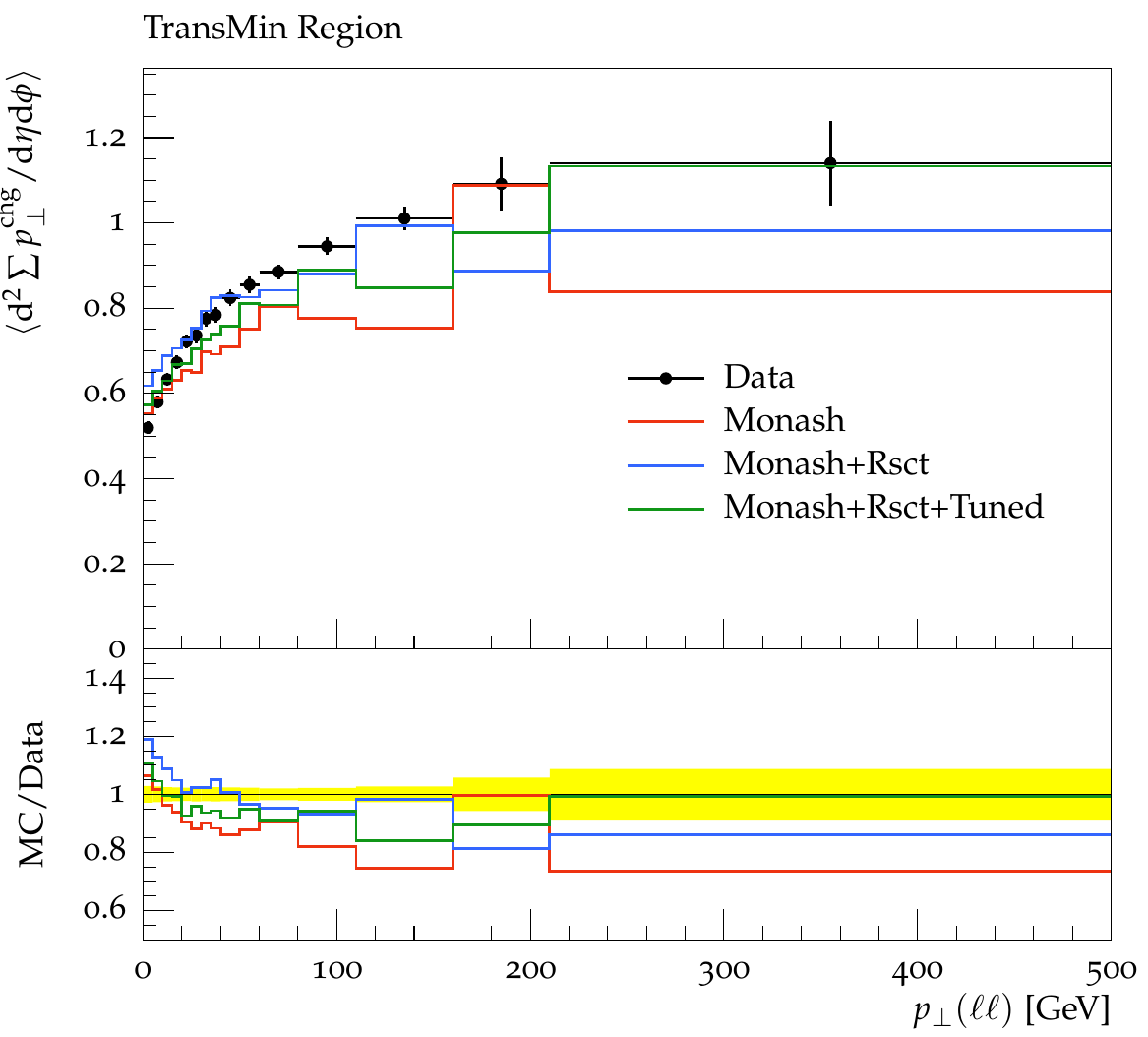}\\[1em]
  \caption{ATLAS data for underlying event distributions in \Z event is compared with \pythiaeight Monash tune without and 
  with rescattering and after tuning prediction for charged particle sum \pT density}
  \label{fig:zue1}
\end{figure}

The next part of the study was the comparison ATLAS UE analysis in \Z events. In \FigRef{fig:zue1}, the 
average charged particle \pT-sum density~\footnote{
The average value in each angular region is divided by the angular area to arrive
at the densities, which allows for direct comparisons between different regions.
} 
is shown in toward and tranmin regions, compared with three different \pythiaeight
predictions. The red line, which is generated with the Monash tune, without any extra
settings, describes the data reasonably well at the low \ZpT\ range. It must be noted, that \pythiaeight
being a leading order PS generator, it is not expected to describe the production of extra jets
from HS, which is occurs more with increasing \ZpT\ . So the focus will be on the 
relatively low \ZpT\ range, $\ZpT < 50$ GeV in this distributions.
When rescattering option is turned on, as in the blue line, it produces too much activity, and the previous
decent agreement between data and simulation is destroyed. To restore the agreement, a simple
tuning of the two MPI parameters is done. It can be seen that with this tuning, the original agreement
of Monash tune can be restored with rescattering. The tuned values of the parameters are:\\
$\kbd{MultipartonInteractions:pT0Ref = 2.2}$\\
$\kbd{MultipartonInteractions:alphaSvalue =0.126}$\\

\begin{figure}[h]
  \centering
  \includegraphics[width=0.45\textwidth]{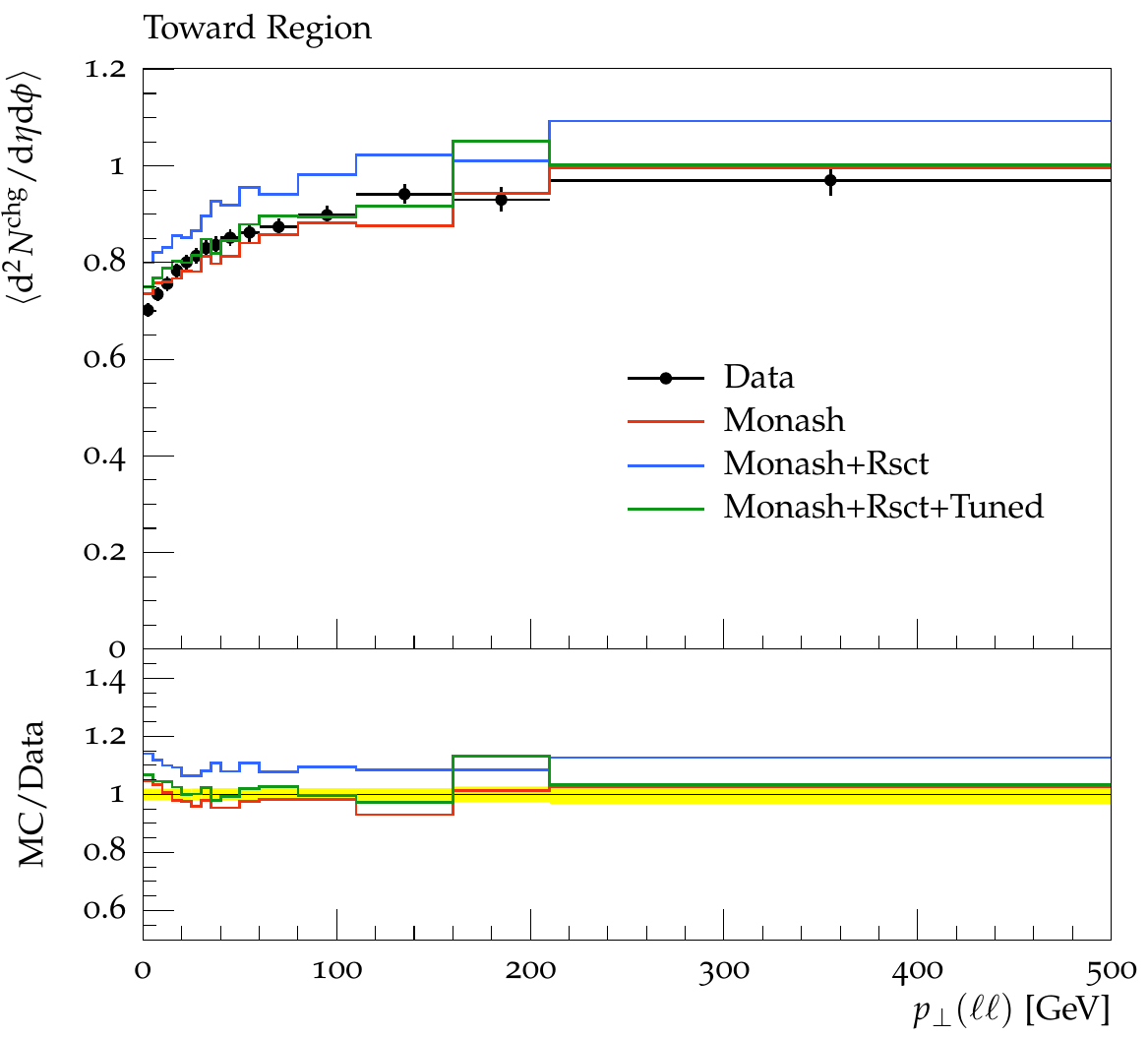}\qquad
  \includegraphics[width=0.45\textwidth]{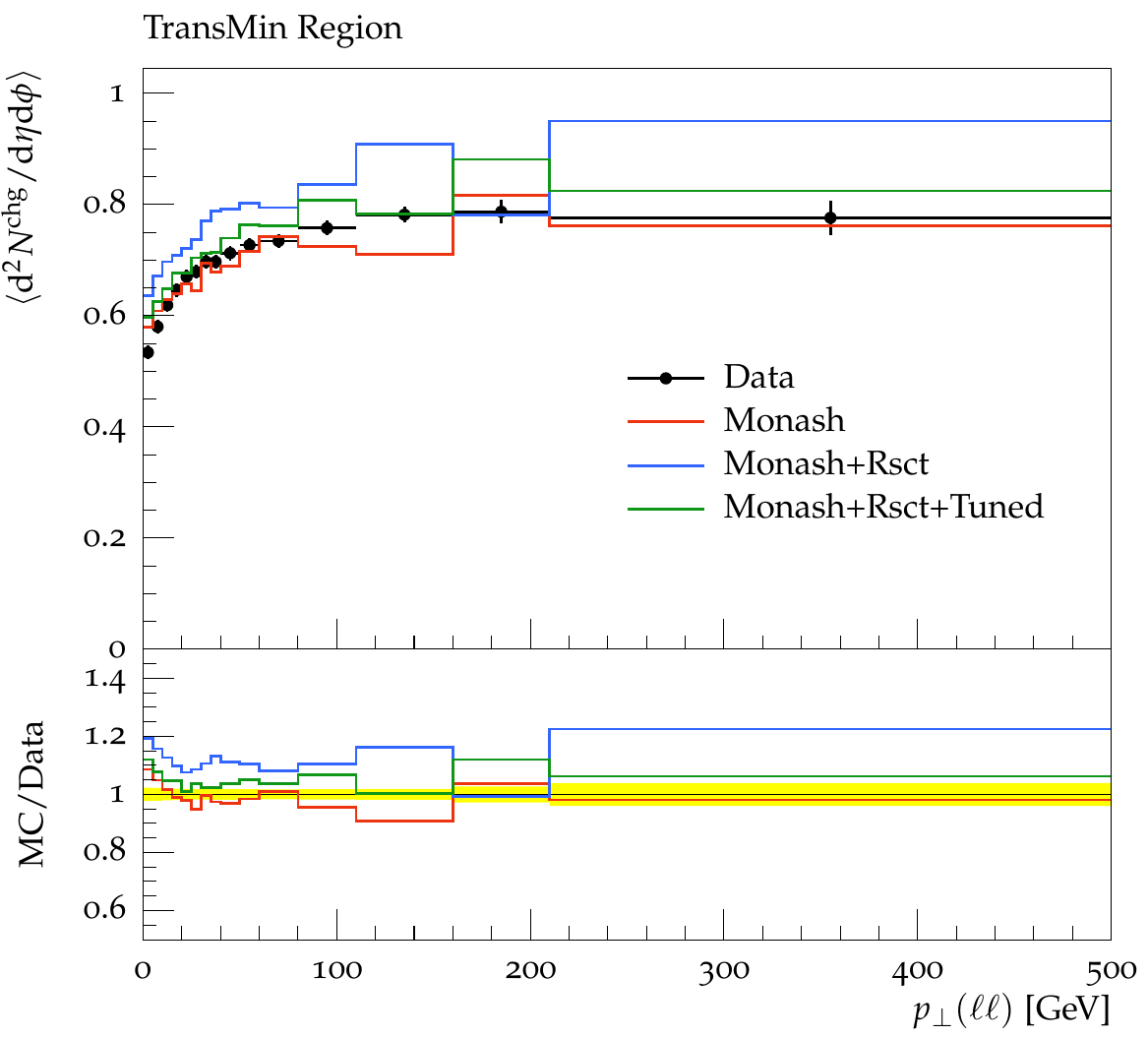}\\[1em]
  \caption{ATLAS data for underlying event distributions in \Z event is compared with \pythiaeight Monash tune without and 
  with rescattering and after tuning prediction for charged particle multiplicity density}
  \label{fig:zue2}
\end{figure}

\begin{figure}[h]
  \centering
  \includegraphics[width=0.45\textwidth]{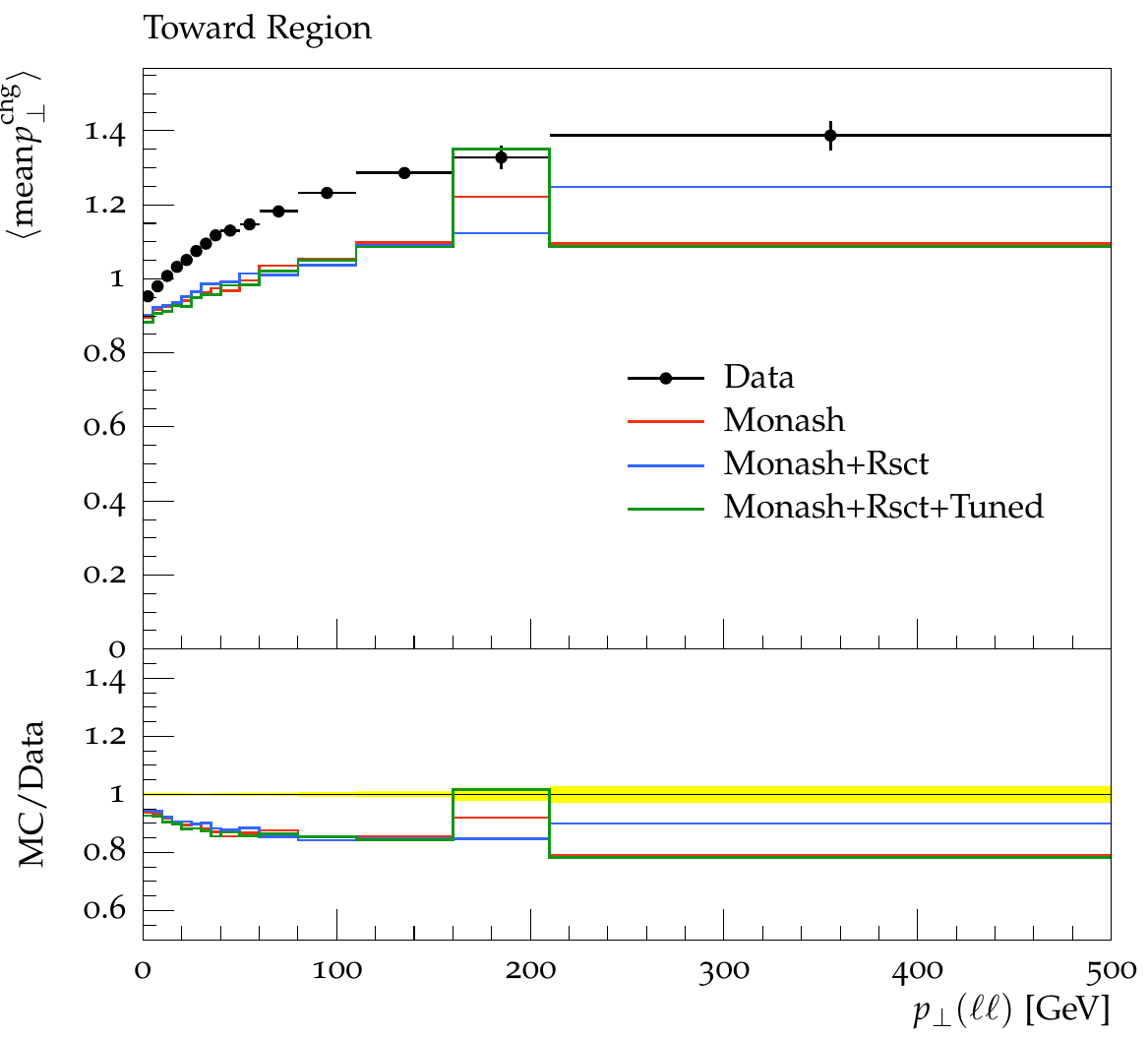}\qquad
  \includegraphics[width=0.45\textwidth]{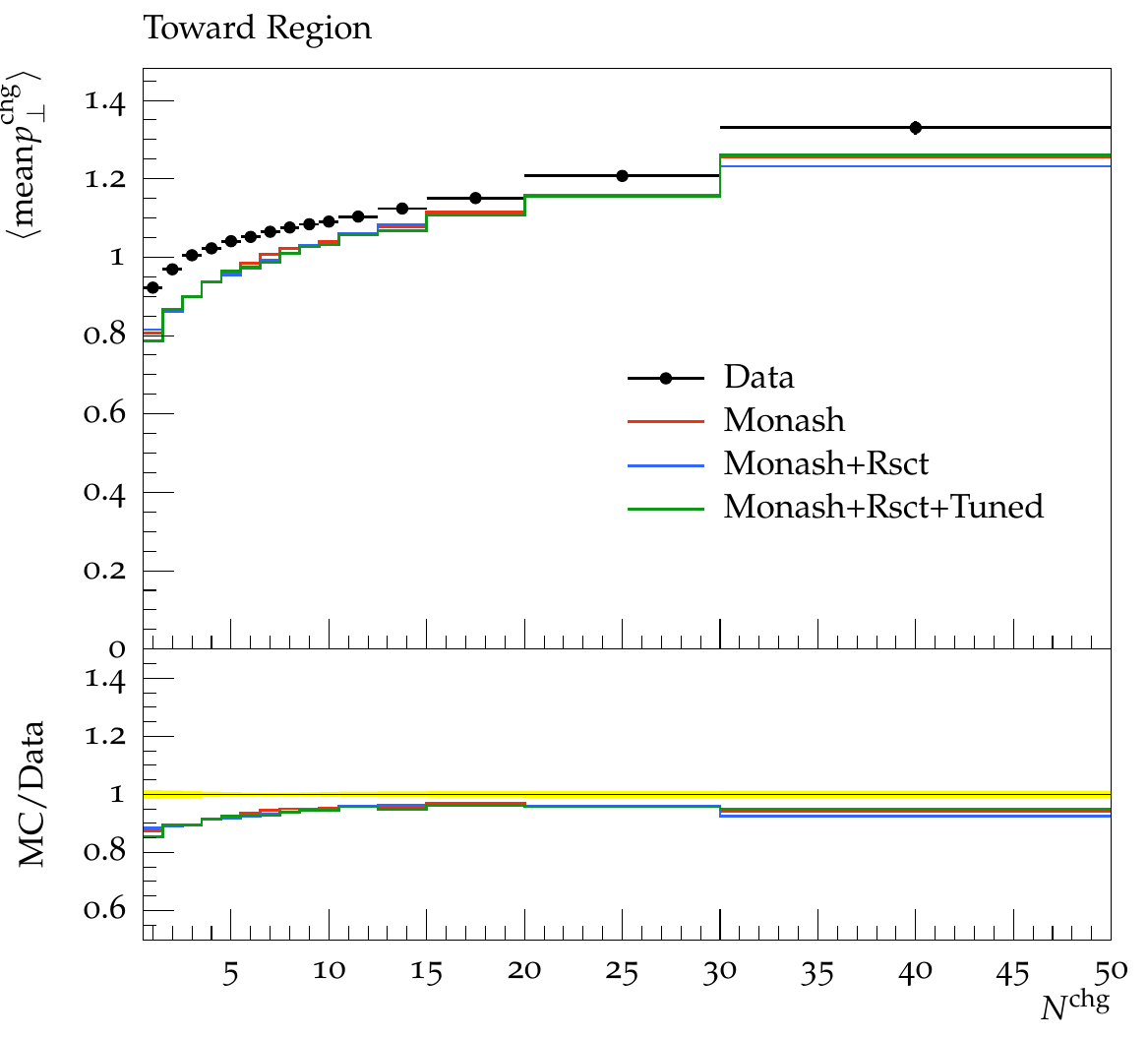}\\[1em]
  \caption{ATLAS data for underlying event distributions in \Z event is compared with \pythiaeight Monash tune without and 
  with rescattering and after tuning prediction for mean charged particle \pT}
  \label{fig:zue3}
\end{figure}

Similar conclusions can be arrived at charged particle multiplicity density shown in \FigRef{fig:zue2}.
In  \FigRef{fig:zue3}, the mean \pT against \Z \pT and multiplicity are not affected at all by rescattering.

\FloatBarrier

%%%%%%%%%%%%%%%%%%%%%%%%%%%%%%%%%%%%%%%%%%%%%%%%%%%
\section{Conclusions}

The effect of rescattering in the framework of multiple parton interactions 
in \pythiaeight are probed. While kinematical observables
constructed from \Z events with 3 additional jets did not
show a strong sensitivity, the charged particle distributions
sensitive to UE is affected by rescattering. A similar level of agreement
with ATLAS \Z UE data can be achieved with rescattering by a simple
tuning of MPI parameters.

%%%%%%%%%%%%%%%%%%%%%%%%%%%%%%%%%%%%%%%%%%%%%%%%%%%
\bibliographystyle{atlasnote}
\bibliography{ref}

\end{document}